\documentclass[trackchanges]{aastex701}

\begin{document}

\title{Magnetoacoustic Portals in Quiet-Sun Fluxtubes Revealed by Chromospheric Spectropolarimetry with SUNRISE \textsc{III}/SCIP
}

\author[0000-0002-7044-6281]{Takayoshi Oba}
\affiliation{Advanced Research Center for Space Science and Technology,
  Institute of Science and Engineering, Kanazawa University,
  Kakuma-machi, Kanazawa, Ishikawa 920-1192, Japan}
\affiliation{Max-Planck-Institut für Sonnensystemforschung, Justus-von-Liebig-Weg 3, 37077 Göttingen, Germany}
\email{oba-takayoshi@staff.kanazawa-u.ac.jp}

\author[orcid=0000-0002-5054-8782,sname='Katsukawa']{Yukio~Katsukawa} \affiliation{National Astronomical Observatory of Japan, 2-21-1 Osawa, Mitaka, Tokyo 181-8588, Japan}\affiliation{Department of Astronomy, The University of Tokyo, 7-3-1, Hongo, Bunkyo-ku, Tokyo 113-0033, Japan}
\affiliation{Department of Astronomical Science, The Graduate University for Advanced Studies (SOKENDAI), 2-21-1 Osawa, Mitaka, Tokyo 181-8588, Japan}
\email{yukio.katsukawa@nao.ac.jp}

\author[orcid=0000-0001-5616-2808,sname='Kubo']{Masahito~Kubo} \affiliation{National Astronomical Observatory of Japan, 2-21-1 Osawa, Mitaka, Tokyo 181-8588, Japan}
\affiliation{Department of Astronomical Science, The Graduate University for Advanced Studies (SOKENDAI), 2-21-1 Osawa, Mitaka, Tokyo 181-8588, Japan}
\email{masahito.kubo@nao.ac.jp}

\author[orcid=0000-0001-7452-0656,sname='Kawabata']{Yusuke~Kawabata} \affiliation{National Astronomical Observatory of Japan, 2-21-1 Osawa, Mitaka, Tokyo 181-8588, Japan}\email{kawabata.yusuke@nao.ac.jp}

\author[orcid=0000-0002-1043-9944,sname='Matsumoto']{Takuma~Matsumoto} \affiliation{Centre for Integrated Data Science, Institute for Space-Earth Environmental Research, Nagoya University, Furocho, Chikusa-ku, Nagoya, Aichi 464-8601, Japan}\email{takuma.matsumoto@gmail.com}

\author[orcid=0000-0002-4669-5376,sname='Ishikawa']{Ryohtaroh~T.~Ishikawa} \affiliation{National Institute for Fusion Science, 322-6 Oroshi-cho, Toki City 509-5292, Japan}\email{ishikawa.ryohtaro@nifs.ac.jp}

\author[orcid=0000-0001-6793-8528,sname='Naito']{Yoshihiro~Naito} \affiliation{Department of Astronomical Science, The Graduate University for Advanced Studies (SOKENDAI), 2-21-1 Osawa, Mitaka, Tokyo 181-8588, Japan}
\affiliation{National Astronomical Observatory of Japan, 2-21-1 Osawa, Mitaka, Tokyo 181-8588, Japan}
\email{yoshihiro.naito@grad.nao.ac.jp}

\author[orcid=0000-0003-4764-6856,sname='Shimizu']{Toshifumi~Shimizu} \affiliation{Department of Earth and Planetary Science, The University of Tokyo, 7-3-1, Hongo, Bunkyo-ku, Tokyo 113-0033, Japan}\affiliation{Institute of Space and Astronautical Science, Japan Aerospace Exploration Agency, 3-1-1, Yoshinodai, Chuo-ku, Sagamihara, Kanagawa 252-5210, Japan}\email{shimizu.toshifumi@isas.jaxa.jp}

\author[orcid=0000-0001-5686-3081,sname='Hara']{Hirohisa~Hara} \affiliation{National Astronomical Observatory of Japan, 2-21-1 Osawa, Mitaka, Tokyo 181-8588, Japan}
\email{hirohisa.hara@nao.ac.jp}

\author[sname='Uraguchi']{Fumihiro~Uraguchi}
\affiliation{National Astronomical Observatory of Japan, 2-21-1 Osawa, Mitaka, Tokyo 181-8588, Japan}
\email{fumihiro.uraguchi@nao.ac.jp}

\author[orcid=0000-0002-8342-8314,sname='Tsuzuki']{Toshihiro~Tsuzuki} \affiliation{National Astronomical Observatory of Japan, 2-21-1 Osawa, Mitaka, Tokyo 181-8588, Japan}
\email{toshihiro.tsuzuki@nao.ac.jp}

\author{Kazuya Shinoda}
\affiliation{National Astronomical Observatory of Japan, 2-21-1 Osawa, Mitaka, Tokyo 181-8588, Japan}
\email{shinoda.kazuya@nao.ac.jp}

\author{Tomonori Tamura}
\affiliation{National Astronomical Observatory of Japan, 2-21-1 Osawa, Mitaka, Tokyo 181-8588, Japan}
\email{tomonori.tamura@nao.ac.jp}

\author[orcid=0000-0003-4452-858X,sname='Suematsu']{Yoshinori Suematsu} \affiliation{National Astronomical Observatory of Japan, 2-21-1 Osawa, Mitaka, Tokyo 181-8588, Japan}\email{yoshinori.suematsu@nao.ac.jp}

\author[orcid=0000-0002-3387-026X,sname='del~Toro~Iniesta']{Jose~Carlos~del~Toro~Iniesta} \affiliation{Instituto de Astrofísica de Andalucía, CSIC, Glorieta de la Astronomía s/n, 18008 Granada, Spain}\affiliation{Spanish Space Solar Physics Consortium}\email{jti@iaa.es}

\author[orcid=0000-0001-8829-1938,sname='Orozco~Suárez']{David~Orozco~Suárez} \affiliation{Instituto de Astrofísica de Andalucía, CSIC, Glorieta de la Astronomía s/n, 18008 Granada, Spain}\affiliation{Spanish Space Solar Physics Consortium}\email{orozco@iaa.es}

\author[orcid=0000-0003-4738-7727,sname='Balaguer~Jiménez']{Maria~Balaguer~Jiménez} \affiliation{Instituto de Astrofísica de Andalucía, CSIC, Glorieta de la Astronomía s/n, 18008 Granada, Spain}\affiliation{Spanish Space Solar Physics Consortium}\email{balaguer@iaa.es}

\author[orcid=0000-0003-0175-6232,sname='Siu-Tapia']{Azaymi~L.~Siu-Tapia} \affiliation{Instituto de Astrofísica de Andalucía, CSIC, Glorieta de la Astronomía s/n, 18008 Granada, Spain}\affiliation{Spanish Space Solar Physics Consortium}\email{siu@iaa.es}

\author[orcid=0000-0001-5518-8782,sname='Quintero Noda']{Carlos~Quintero Noda} \affiliation{Instituto de Astrofísica de Canarias, Vía Láctea, s/n, E-38205 La Laguna, Spain}\affiliation{Universidad de La Laguna, E-38205 La Laguna, Spain}\affiliation{Spanish Space Solar Physics Consortium}\email{carlos.quintero@iac.es}

\author[orcid=0000-0002-3418-8449,sname='Solanki']{Sami~K.~Solanki} \affiliation{Max-Planck-Institut für Sonnensystemforschung, Justus-von-Liebig-Weg 3, 37077 Göttingen, Germany}\email{solanki@mps.mpg.de}

\author[orcid=0000-0003-1459-7074,sname='Lagg']{Andreas~Lagg} \affiliation{Max-Planck-Institut für Sonnensystemforschung, Justus-von-Liebig-Weg 3, 37077 Göttingen, Germany}\email{lagg@mps.mpg.de}

\author[orcid=0000-0002-9972-9840,sname='Gandorfer']{Achim~Gandorfer} \affiliation{Max-Planck-Institut für Sonnensystemforschung, Justus-von-Liebig-Weg 3, 37077 Göttingen, Germany}\email{gandorfer@mps.mpg.de}

\author[orcid=0000-0002-0787-8954,sname='Bernasconi']{Pietro~Bernasconi} \affiliation{Johns Hopkins University Applied Physics Laboratory, 11100 Johns Hopkins Road, Laurel, Maryland, USA}\email{pietro.bernasconi@jhuapl.edu}

\author[sname='Berkefeld']{Thomas~Berkefeld} \affiliation{Institut für Sonnenphysik (KIS), Georges-Köhler-Allee 401a, 79110 Freiburg, Germany}\email{thomas.berkefeld@leibniz-kis.de}

\author[orcid=0009-0009-4425-599X,sname='Feller']{Alex~Feller} \affiliation{Max-Planck-Institut für Sonnensystemforschung, Justus-von-Liebig-Weg 3, 37077 Göttingen, Germany}\email{feller@mps.mpg.de}

\author[orcid=0000-0001-6317-4380,sname='Riethmüller']{Tino~L.~Riethmüller} \affiliation{Max-Planck-Institut für Sonnensystemforschung, Justus-von-Liebig-Weg 3, 37077 Göttingen, Germany}\email{riethmueller@mps.mpg.de}

\author[orcid=0000-0001-9228-3412,sname='Álvarez-Herrero']{Alberto~Álvarez-Herrero} \affiliation{Instituto Nacional de T\'ecnica Aeroespacial (INTA), Ctra. de Ajalvir, km. 4, E-28850 Torrejón de Ardoz, Spain}\affiliation{Spanish Space Solar Physics Consortium}\email{alvareza@inta.es}

\author[orcid=0000-0003-3490-6532,sname='Smitha']{H.~N.~Smitha} \affiliation{Max-Planck-Institut für Sonnensystemforschung, Justus-von-Liebig-Weg 3, 37077 Göttingen, Germany}\email{narayanamurthy@mps.mpg.de}

\author[sname='Grauf']{Bianca~Grauf} \affiliation{Max-Planck-Institut für Sonnensystemforschung, Justus-von-Liebig-Weg 3, 37077 Göttingen, Germany}\email{grauf@mps.mpg.de}

\author[sname='Carpenter']{Michael~Carpenter} \affiliation{Johns Hopkins University Applied Physics Laboratory, 11100 Johns Hopkins Road, Laurel, Maryland, USA}\email{michael.carpenter@jhuapl.edu}

\author[sname='Bell']{Alexander~Bell} \affiliation{Institut für Sonnenphysik (KIS), Georges-Köhler-Allee 401a, 79110 Freiburg, Germany}\email{albe@leibniz-kis.de}

\author[orcid=0000-0001-7764-6895,sname='Martínez~Pillet']{Valentín~Martínez~Pillet} \affiliation{Instituto de Astrofísica de Canarias, Vía Láctea, s/n, E-38205 La Laguna, Spain}\affiliation{Spanish Space Solar Physics Consortium}\email{vmpillet@iac.es}

\author[orcid=0000-0002-7318-3536,sname='Bailén']{Francisco~Javier~Bailén} \affiliation{Instituto de Astrofísica de Andalucía, CSIC, Glorieta de la Astronomía s/n, 18008 Granada, Spain}\affiliation{Spanish Space Solar Physics Consortium}\email{fbailen@iaa.es}

\author[orcid=0000-0002-2055-441X,sname='Blanco~Rodríguez']{Julian~Blanco~Rodríguez} \affiliation{Universitat de Valencia Catedrático José Beltrán 2, E-46980 Paterna-Valencia, Spain}\affiliation{Spanish Space Solar Physics Consortium}\email{julian.blanco@uv.es}

\author[orcid=0000-0003-4319-2009,sname='Castellanos~Durán']{Juan~Sebastián~Castellanos~Durán} \affiliation{Max-Planck-Institut für Sonnensystemforschung, Justus-von-Liebig-Weg 3, 37077 Göttingen, Germany}\email{castellanos@mps.mpg.de}

\author[orcid=0009-0002-6808-5154,sname='Harnes']{Edvarda~Harnes} \affiliation{Max-Planck-Institut für Sonnensystemforschung, Justus-von-Liebig-Weg 3, 37077 Göttingen, Germany}\email{harnes@mps.mpg.de}

\author[orcid=0000-0001-6029-7529,sname='Hölken']{Johannes~Hölken} \affiliation{Max-Planck-Institut für Sonnensystemforschung, Justus-von-Liebig-Weg 3, 37077 Göttingen, Germany}\email{hoelken@mps.mpg.de}

\author[orcid=0000-0003-1409-1145,sname='Iglesias']{Francisco~A.~Iglesias} \affiliation{Max-Planck-Institut für Sonnensystemforschung, Justus-von-Liebig-Weg 3, 37077 Göttingen, Germany}\affiliation{Grupo de Estudios en Heliofísica de Mendoza, CONICET, Universidad de Mendoza, Boulogne sur Mer 683, 5500 Mendoza, Argentina}\email{iglesias@mps.mpg.de}

\author[orcid=0000-0003-1483-4535,sname='Strecker']{Hanna~Strecker} \affiliation{Instituto de Astrofísica de Andalucía, CSIC, Glorieta de la Astronomía s/n, 18008 Granada, Spain}\affiliation{Spanish Space Solar Physics Consortium}\email{streckerh@iaa.es}

\author[orcid=0000-0003-1971-5551,sname='Vukadinović']{Dušan~Vukadinović} \affiliation{Institut für Physik, Universität Graz, Universitätsplatz 5, 8010 Graz, Austria}\affiliation{Max-Planck-Institut für Sonnensystemforschung, Justus-von-Liebig-Weg 3, 37077 Göttingen, Germany}\email{dusan.vukadinovic@uni-graz.at}

\begin{abstract}
Acoustic waves propagate into the chromosphere, contributing to energy transport and their dynamics. 
Their upward propagation is restricted to frequencies above the acoustic cutoff frequency. 
The magnetic field configuration plays a key role in determining whether acoustic waves can propagate upward because the cutoff frequency is reduced in regions where the field is inclined with respect to gravity, forming so-called magnetoacoustic portals. 
Previous studies linked magnetic fields and oscillations in quiet regions, but these analyses were based on photospheric magnetic field information, leaving chromospheric structure unconstrained. 
This study investigates the coupling between acoustic waves and magnetic topology using photospheric and, for the first time, chromospheric spectropolarimetry in a quiet region, obtained with the Sunrise Chromospheric Infrared SpectroPolarimeter (SCIP) aboard the {\sc Sunrise~iii} balloon-borne solar observatory launched in 2024. 
The SCIP sit-and-stare observations sampled magnetic features in which the line-of-sight field strength exhibits multiple sharp spatial peaks in the photosphere while becoming broader and weaker at two heights in the chromosphere, indicating expanding fluxtubes. 
The chromospheric velocity field in these fluxtubes exhibits strong 5-minute oscillations, while the surrounding regions show weak 3-minute oscillations. 
In these fluxtubes, sawtooth temporal velocity variations are associated with intensity enhancements, suggesting steepened shocks. 
Fluxtubes with low-frequency oscillations are identified not only in network regions but also in weak internetwork regions. 
These results provide observational evidence that fluxtubes expanding into the chromosphere act as magnetoacoustic portals, in both network and internetwork regions, allowing low-frequency waves to propagate upward and driving chromospheric dynamics via shocks. 
\end{abstract}



\section{Introduction}
Acoustic waves are observed ubiquitously on the solar surface, where their power is largely concentrated at a timescale of about 5 minutes \citep{Ulrich1970}. 
Acoustic waves carry a large amount of kinetic energy, but it is still controversial how much they contribute to energy input for heating and drive dynamics in the upper atmosphere. 
Most of the acoustic wave energy lies at frequencies lower than the cutoff frequency due to atmospheric stratification and is trapped in the photosphere, where waves become evanescent and cannot propagate to higher layers. 
\cite{Fossum2005} and \citet{Carlsson2007} estimated the total energy flux in high-frequency acoustic waves and found it to be insufficient to balance the radiative losses in the chromosphere \citep{Vernazza1981}. 
On the other hand, high spatial-resolution observations have revealed a much larger acoustic energy flux \citep{Bello_Gonzalez2010}, comparable to the chromospheric energy loss. \\ \indent
The acoustic cutoff frequency plays a crucial role in determining which acoustic waves can travel from the photosphere to the chromosphere.
In non-magnetized regions, the cutoff frequency is estimated to be about 5.2 mHz in an isothermal atmosphere \citep{Felipe2020}, corresponding to a period of approximately 3 min. 
Consequently, the dominant 5-minute oscillatory components (3 mHz) lie below the cutoff frequency and are therefore trapped in the photosphere. 
However, the cutoff frequency varies depending on atmospheric conditions. 
One possible factor is radiative damping: 
energy exchange through radiation leads to a non-adiabatic response to the perturbation and weakens the restoring force, thereby lowering the cutoff frequency \citep{Centeno2006, Centeno2009b, Khomenko2008, Felipe2020}. 
Another factor affecting the cutoff frequency is the magnetic field structure. 
A theoretical study by \citealt{Bel1977} demonstrated that when the plasma $\beta$ (defined as the ratio of gas to magnetic pressure) is low, the cutoff frequency is reduced by a factor of \textrm{cos$\theta$}, where $\theta$ is the inclination of the magnetic field with respect to the direction of gravity. 
This is commonly referred to as a \textit{magnetoacoustic portal}. 
Observational studies in active regions have largely confirmed this theoretical prediction. 
The penumbra provides a clear example of an environment with inclined magnetic fields, and has been observed to exhibit a decrease in the cutoff frequency with increasing field inclination \citep{McIntosh2006}. 
\citet{Koyama2024} reported that a specific range of inclination of magnetic field in a plage contributes to heating the chromosphere. 
In quiet regions, \citet{Jefferies2006} showed that the spatial power spectrum derived from travel-time analysis peaks at the supergranulation-scale. 
This result indicates that network magnetic field structures act as magnetoacoustic portals, channeling low-frequency acoustic waves upward. 
Taking this effect into account, the energy flux carried by low-frequency acoustic waves ($<$5.2 mHz) is estimated to be a factor of 4 or more greater than that carried by high-frequency waves ($>$5.2 mHz), amounting to several tens of percent or even up to half of the energy required to maintain chromospheric heating \citep{Jefferies2006, Rajaguru2019}. \\ \indent
The magnetic topology is therefore a key factor in determining how acoustic waves propagate into the chromosphere, as chromospheric oscillations are observed to be strongly modulated by network magnetic structures \citep{Lites1993, Deubner1990}. 
However, previous studies relied on photospheric magnetic field measurements or extrapolations from photospheric boundary conditions. 
This approach is inherently limited for investigating magnetoacoustic portals because the acoustic cutoff frequency varies with atmospheric height. 
Observational studies suggest that the cutoff frequency increases toward the chromosphere \citep{Wisniewska2016}, while atmospheric models predict that it reaches a maximum near or above the temperature minimum layer \citep{Felipe2020}. 
Therefore, direct measurements of the magnetic field from the photosphere to the chromosphere are required to understand how acoustic waves propagate through magnetoacoustic portals \citep{Sangal2024}.  \\ \indent
To overcome this limitation, SCIP onboard Sunrise offers accurate spectropolarimetric observations that allow the magnetic field to be measured from the photosphere and the chromosphere.
This study, for the first time, reports the propagation properties of acoustic waves toward the chromosphere based on simultaneous measurements of the magnetic field in both the photosphere and chromosphere, using the Sunrise Chromospheric Infrared Spectropolarimeter (SCIP: \citealt{Katsukawa2026}), onboard the stratospheric balloon-borne observatory ({\sc Sunrise~iii}: \citealt{Korpi-Lagg2025}). 

\section{Observations} \label{sec:style}
\begin{figure}[h!]
   \centering
   \includegraphics[width=1.0\linewidth]{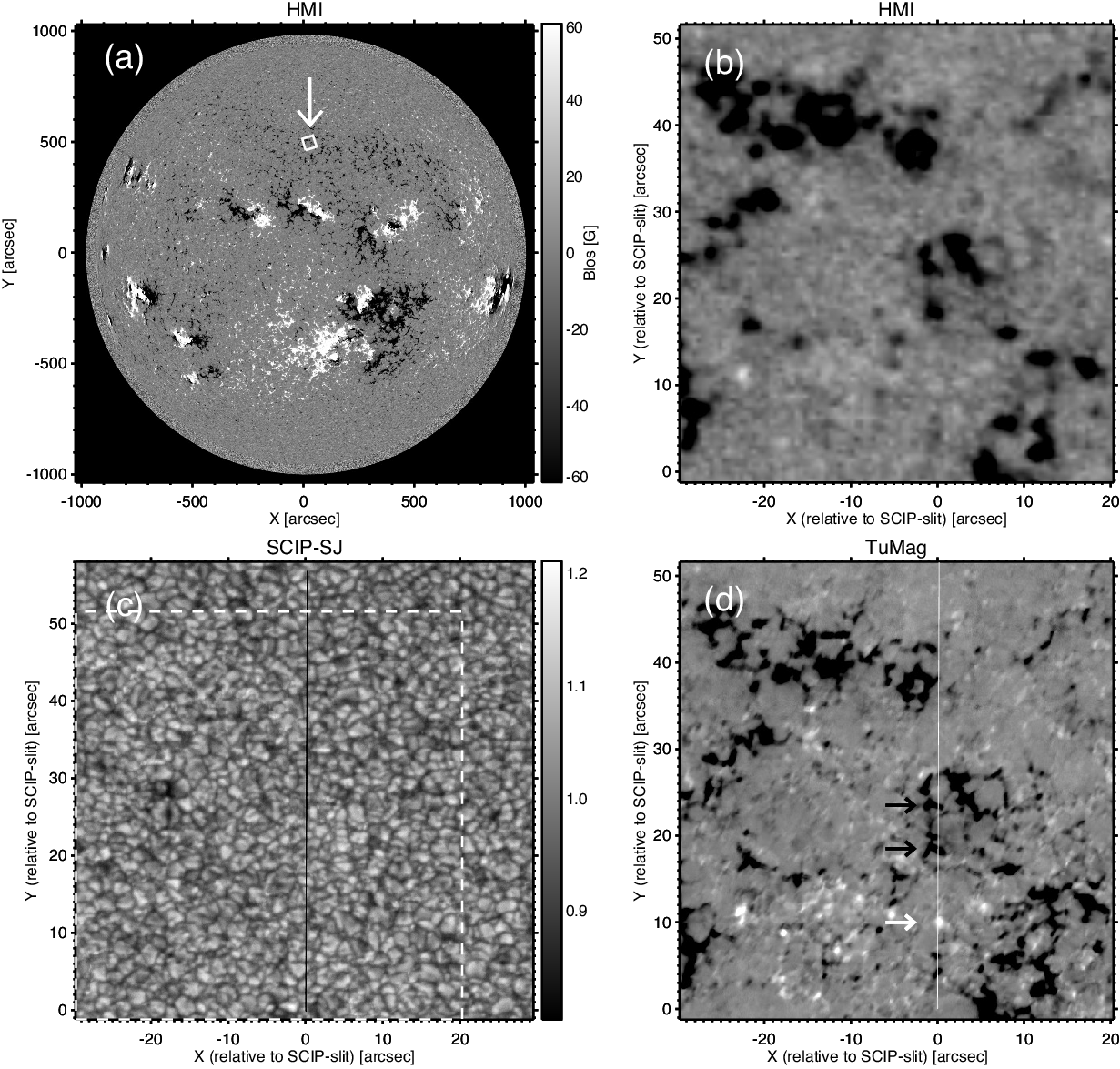}
      \caption{
      Panel (a): LOS magnetogram taken by SDO/HMI at 03:41:08 on 16th July 2024. 
      The white square outlines the region covered by TuMag. 
      The color scale is shared with the other magnetograms shown in panels (b) and (d). 
      Panel (b): Enlarged view of the LOS magnetogram in panel (a), rotated clockwise by 17.1 degrees so that the axes are aligned with those of SCIP and TuMag. 
      The coordinates are relative values with respect to the southern end of the SCIP-slit. 
      Panel (c): SCIP-SJ image acquired at 03:40:52.537. 
      The dashed white rectangular box indicates the overlapped region of the TuMag magnetogram (before phase-diversity reconstruction),  corresponding to the full field-of-view in panels (b) and (d). 
      Panel (d): Coaligned LOS magnetograph by TuMag. 
      The vertical white line is overlaid for reference to mark the SCIP slit position, inferred by comparing the SCIP-SJ image in panel (c) with a TuMag intensity image (not shown). 
\label{fig:cc}}
         \label{fig1}
   \end{figure}

{\sc Sunrise} is a balloon-borne solar observatory equipped with a 1-meter aperture telescope \citep{Barthol2011, Solanki2010, Solanki2017}. 
The latest and strongly updated version, {\sc Sunrise iii} \citep{Korpi-Lagg2025}, was successfully launched on 10 July 2024, and observations were carried out at stratospheric altitudes, providing seeing-free conditions \citep{Solanki2026}. 
Unprecedented pointing stability was achieved through a new gondola \citep{Bernasconi2025}, in combination with the Correlating Wavefront Sensor (CWS: \citealt{Berkefeld2026}). \\ \indent
This study uses the spectropolarimetric data obtained with SCIP, one of the three science instruments onboard the balloon-borne observatory. 
SCIP is a slit spectropolarimeter (SP) that simultaneously observes two near-infrared bands at 850 and 770 nm. 
These wavelength bands contain numerous spectral lines, thereby seamlessly covering heights ranging from the photosphere to the chromosphere \citep{Quintero_Noda2017, Quintero_Noda2017b, Quintero_Noda2018, Matsumoto2023, Kawabata2024}. 
In addition to the two spectropolarimetric channels, SCIP is equipped with a slit-jaw (SJ) camera that provides two-dimensional continuum intensity maps, complementing the one-dimensional slit observations. 
The detailed SCIP specifications are summarized in Table \ref{tbl:inst}. 
The SP and SJ data were processed using standard data reduction and calibration procedures \citep{Solanki2026}. \\ \indent
This work uses a dataset obtained in the fixed slit position (sit-and-stare) mode, which is well suited for investigating the dynamical behavior of the chromosphere \citep{Mathur2022}. 
The observations were carried out on 16 July 2024 from 03:40:52 to 04:04:30 UT, with a total duration of 23 min 38 sec.
Stokes $I, Q, U,$ and $V$ were recorded at each integration, at a cadence of 1.536 sec. 
Note that the observations were carried out at a heliocentric angle of 29.5 degrees corresponding to $\mu=0.87$; nevertheless, the line-of-sight (LOS) predominantly reflects the vertical component. \\ \indent
The observations targeted a quiet region, as demonstrated by the SDO/HMI magnetogram \citep{Pesnell2012, Scherrer2012} in Fig.\ref{fig1}(a) and its enlarged view in Fig.\ref{fig1}(b). 
The spatial context of magnetic structures cannot be provided from the SCIP sit-and-stare observations alone. 
This limitation is mitigated by the Tunable Magnetograph (TuMag; \citealt{delToroIniesta2025}) onboard {\sc Sunrise III}, which provides two-dimensional magnetograms at visible wavelengths. 
Based on the alignment between the SCIP-SJ image (Fig.\ref{fig1}c) and the TuMag intensity image (not shown), the SCIP slit position is overlaid for reference on the LOS magnetic field strength map in Fig.\ref{fig1}(d), derived from the \ion{Fe}{1} 5250.2\AA~line using the weak-field approximation (WFA; \citealt{Jefferies1989}).  
The SCIP slit samples several magnetic patches of negative polarity, for example at $y=18^{\prime\prime}$ and $23^{\prime\prime}$ (black arrows). 
The SDO/HMI magnetogram shown in Fig.~\ref{fig1}~(b) exhibits corresponding magnetic features with reduced contrast, suggesting that these structures are associated with the magnetic network. 
A weak positive-polarity patch is identified at $y=10^{\prime\prime}$ (white arrow) in the TuMag magnetogram, likely representing an internetwork magnetic element. 
It is barely detectable in the HMI magnetogram, suggesting a spatial scale close to the HMI resolution limit ($\sim1^{\prime\prime}$), comparable to the scale at which internetwork magnetic elements start to be resolved \citep{BellotRubio2019}. 

\begin{table}
\begin{center}
\caption{Specification of the SCIP}
\label{tbl:inst}
\begin{tabular}{lr}
\hline \hline
\multicolumn{2}{l}{\textbf{SP (850nm band)}} \\ \hline
Spatial resolution & 0.21$^{\prime\prime}$ \\
Spectral resolution & 37.93 m\AA \\
Slit length & 58$^{\prime\prime}$ \\
Plate scale & 0.0936$^{\prime\prime}$ \\
Wavelength coverage & 8463.7000-8548.5855 \AA \\
Spectral sampling & 39.5 m\AA \\
\hline
\multicolumn{2}{l}{\textbf{SJ}} \\ \hline
Wavelength band & 7702.08-7712.08 \AA \\
Plate scale & 0.0936$^{\prime\prime}$ \\
Field of View & 60$^{\prime\prime} \times 60^{\prime\prime}$ \\
Exposure time & 10 msec \\
\hline \hline
\end{tabular}
\end{center}
\end{table}

\section{Methodology} \label{sec:floats}
While SCIP covers many spectral lines, this study focuses on a subset of the 850nm band: 
the photospheric \ion{Fe}{1} 8468\AA~line, with a Land\'e factor of $g=2.50$, and the chromospheric \ion{Ca}{2} 8498\AA~and \ion{Ca}{2} 8542\AA~lines, with effective Land\'e factors of $g_{\mathrm{eff}}=1.07$ and $g_{\mathrm{eff}}=1.10$, respectively. 
The chromospheric \ion{Ca}{2} 8542\AA~line is formed higher than the \ion{Ca}{2} 8498\AA~line \citep{Quintero_Noda2017}. \\ \indent 
The chromospheric line-core intensity is defined as the minimum intensity of the \ion{Ca}{2} 8542\AA~line, at each spatial and temporal pixel, denoted as $I_{\mathrm{Ca\,II}\, 8542}$. 
The intensity is normalized to the continuum intensity in the wavelength range 8507.150 - 8509.125~\AA, averaged over the full field-of-view and the entire observing period. \\ \indent
Doppler velocities in the photosphere and chromosphere are calculated using the center-of-gravity (COG) method \citep{Uitenbroek2003}, and are denoted as $v_{\mathrm{Fe\,I} \,8468}$ and $v_{\mathrm{Ca\,II}\,8542}$, respectively. 
The wavelength spans used for the COG calculations are $\pm 158$ $\textrm{m\AA}$ (9 pixels) for $v_{\mathrm{Fe\,I}\,8468}$ and $\pm 316$ $\textrm{m\AA} $ (17 pixels) for $v_{\mathrm{Ca\,II} \,8542}$. 
The velocity zero-point was defined by applying a constant offset such that the spatially and temporally averaged Doppler velocity is 0 km/s. 
To isolate the acoustic waves component, a sub-sonic filter \citep{Title1989, Oba2017a} is applied to the two-dimensional dataset covering time and space. 
Acoustic waves are identified as components with phase speeds exceeding 7 km/s \citep{Cauzzi2009}, comparable to the sound speed in the photosphere and the lower chromosphere \citep{Schmitz1998}. \\ \indent
The LOS component of the magnetic field is calculated using the WFA \citep{Centeno2018} for \ion{Ca}{2} 8542 \AA \ and \ion{Ca}{2} 8498 \AA \ and \ion{Fe}{1} 8468 \AA, denoted as $B_{\mathrm{Ca\,II}\,8542}$, $B_{\mathrm{Ca\,II}\,8498}$, and $B_{\mathrm{Fe\,I}\,8468}$. 
The WFA assumes a simplified atmosphere with constant velocity and magnetic field along the LOS, and is applicable provided that the Zeeman splitting is smaller than the spectral line width.
The WFA is applied using the same wavelength windows as those adopted for the Doppler velocity calculations. 
The wavelength window is shifted at each pixel so that it is centered on the Stokes~$I$ line core (defined as the minimum intensity), thereby reducing errors arising from local Doppler shifts.
For comparison, the COG method \citep{Rees1979}, based on the wavelength separation between the COGs of the $I+V$ and $I-V$ profiles, was applied to the photospheric \ion{Fe}{1} 8468 \AA\ line, yielding results consistent with those from the WFA, with deviations below 10\%. 
However, the COG method resulted in larger random errors, reducing its reliability in weak-field regions. 
Therefore, only the WFA results are considered in this study.

\section{Results} \label{sec:displaymath}
In the following sections, the wave properties are compared among representative magnetic environments sampled by the SCIP slit: network magnetic features, internetwork magnetic features, and the weakly magnetized atmosphere. 
 The magnetic configurations are first described in Section 4.1, followed by an examination of their oscillatory properties in Sections 4.2--4.4. 

\subsection{Spatial distribution of magnetic field}
\begin{figure}[h!]
   \centering
   \includegraphics[width=0.79\linewidth]{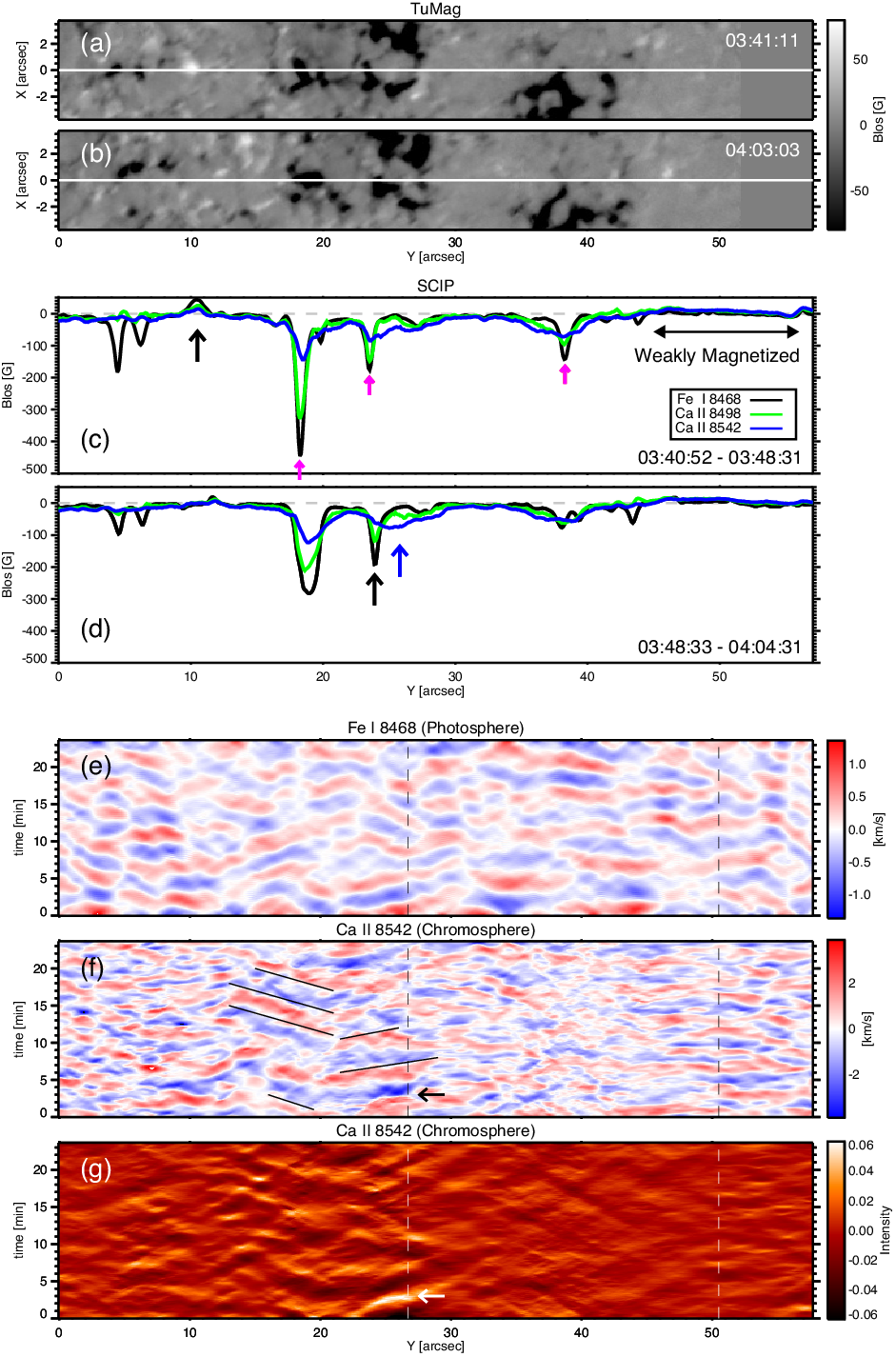}
      \caption{
      Panel (a) and (b): 
      Two-dimensional distributions of the LOS magnetic field strength observed by TuMag at 03:41:11 and 04:03:03, respectively. 
      The white horizontal line indicates the SCIP slit position. 
	  The uniform gray region on the right corresponds to areas not covered by the TuMag observations, where zero values are padded for display purposes. 
	  Panels (c) and (d): LOS magnetic field strength along the SCIP slit for the intervals 03:40:52-03:48:31 and 03:48:33-04:04:31, respectively. 
	  $B_{\mathrm{Fe\,I}\,8468}$, $B_{\mathrm{Ca\,II}\,8498}$, and $B_{\mathrm{Ca\,II}\,8542}$ are shown in black, green, and blue, respectively. 
          The grey dashed line indicates 0 G. 
      Panel (e) and (f): $y-t$ diagrams of $v_{\mathrm{Fe\,I}\,8468}$ and $v_{\mathrm{Ca\,II}\,8542}$. 
	  Panel (g): $y-t$ diagram of $I_{\mathrm{Ca\,II}\,8542}$. 
      The quantities in panels (e)-(g) show components with phase speeds exceeding $7$ km/s. 
      The two dashed lines in panels (e)-(g) indicate the locations from which the $\lambda$-$t$ diagrams in Fig.~\ref{fig3} were extracted.
\label{fig:cc}}
         \label{fig2}
   \end{figure}

Figure~\ref{fig2}(a) and (b) show the LOS component of the magnetic field ($B_{\mathrm{LOS}}$) observed with TuMag near the beginning and end of the observing sequence, at 03:41:11 and 04:03:33, respectively. 
Figure~\ref{fig2} (c) and (d) exhibit $B_{\mathrm{LOS}}$ with SCIP through the WFA applied to the Stokes~$V$ profiles integrated over an early phase (03:40:52-03:48:31) and a later phase (03:48:33-04:04:31) of the observations, respectively.
The SCIP slit intersects several magnetic features with photospheric field strengths of a few hundred Gauss. 
Compared with their photospheric counterparts, the chromospheric magnetic features are weaker and exhibit broader spatial distributions. 
This indicates that the fluxtubes expand with height owing to the atmospheric stratification \citep{Solanki1990, Buente1993, Ishikawa2021}, consistent with the increasing formation heights of \ion{Fe}{1} 8468\AA, \ion{Ca}{2} 8498\AA, and \ion{Ca}{2} 8542\AA~\citep{Quintero_Noda2017b}. 
Throughout this paper, the term \textit{fluxtube} refers to a localized magnetic concentration extending from the photosphere into the chromosphere, without implying a thin cylindrical geometry.
For a rough estimate of the expansion of the fluxtubes, we introduce the expansion factor $\Gamma$, defined here as the ratio of the maximum $B_{\mathrm{Fe \,I} \,8468}$ to the maximum $B_{\mathrm{Ca \,II \,8542}}$ for each fluxtube, rather than the cross-sectional area ratio. 
This quantity should be regarded only as a simple proxy for fluxtube expansion, as the sit-and-stare observations limit this estimate because only a one-dimensional cut through each fluxtube is sampled.
The values of $\Gamma$ for the three negative polarity fluxtubes, located at $y=18^{\prime\prime}, 24^{\prime\prime}$, and $38^{\prime\prime}$ (indicated by small magenta arrows), are 3.1, 2.1, and 2.0, respectively.
A weak positive-polarity fluxtube located at $y=10.5^{\prime\prime}$ (upward arrow) in panel (c), with $B_{\mathrm{Fe \,I} \,8468}=43$ G, also extends into the chromosphere, with an expansion factor of $\Gamma=2.4$. 
A time series of TuMag magnetograms (not shown here) indicates that this fluxtube is present only during the early phase of the observations, and subsequently fragments and disperses, and is therefore no longer visible in panel (b).
It should be noted that, for the magnetic feature at $y=24^{\prime\prime}$ (black arrow) in panel (d), the centroid of the chromospheric magnetic field is significantly displaced from that of the photospheric field, appearing near $y=27^{\prime\prime}$ (blue arrow). 
This displacement may suggest a tilt of the fluxtube axis from the photosphere to the chromosphere. 
However, the two magnetic features at $y=24^{\prime\prime}$ and $y=28^{\prime\prime}$ in the photosphere merge into a single fluxtube in the chromosphere, complicating the interpretation of the centroid displacement. 
Furthermore, the SCIP slit also samples a very weak field region from $y=45^{\prime\prime}$-$58^{\prime\prime}$, where $| B_{\mathrm{LOS}} | < 10 \ \mathrm{G}$ in both the photosphere and chromosphere, indicated by a double-headed arrow. 
This region is referred to as the \textit{weakly-magnetized atmosphere} throughout this paper. 

\subsection{Doppler Velocity}
Figure~\ref{fig2} (e) and (f) show space-time ($y-t$) diagrams of $v_{\mathrm{Fe\,I}\,8468}$ and $v_{\mathrm{Ca\,II}\,8542}$, respectively, with phase speeds exceeding 7 km/s. 
The photospheric velocity pattern is characterized by spatial scales of typically a few to ten arcseconds and temporal scales of a period of 5 min. 
Both spatial and temporal scales are largely uniform and show little dependence on the presence of magnetic fluxtubes \citep{Deubner1990}. 
The chromospheric velocity pattern, however, exhibits a dependence on the local magnetic environment. 
In the weakly-magnetized atmosphere (double-headed arrow in panel~(c)), the dominant period is approximately 3 min, inferred from roughly 8 oscillation cycles over the 24 min observing sequence. 
This behavior indicates the operation of the acoustic cutoff, which suppresses lower-frequency waves and allows only higher-frequency waves to propagate into the chromosphere. 
In contrast, the chromospheric velocity pattern is clearly modulated by the presence of magnetic fluxtubes, particularly at $y=15^{\prime\prime}$-$28^{\prime\prime}$, where the dominant period of the chromospheric velocity fluctuation is approximately 5 min, inferred from roughly 5 oscillation cycles over the 24 min observing sequence. 
Furthermore, the velocity amplitude is significantly enhanced within this fluxtube, with a standard deviation of 2.0 km/s, compared to 0.90 km/s in the weakly magnetized atmosphere. 
The energy flux of acoustic waves is expected to be approximately 5 times larger in the expanding fluxtube, assuming the same plasma density and sound speed. \\ \indent
At $y=7^{\prime\prime}$-$13^{\prime\prime}$, low-frequency oscillations are present only during the early phase of the observations, corresponding to the location of a weak magnetic fluxtube (Section 4.1). 
The disappearance of the low-frequency oscillations coincides with the fragmentation and dispersal of the magnetic feature. 
Changes in the transmission of low-frequency acoustic waves reflect the modification of magnetic structure, as anticipated in \citet{Jefferies2019}. \\ \indent
Another notable feature is the apparent lateral propagation of chromospheric oscillations originating from a localized point within the expanding fluxtube at $y=21^{\prime\prime}$, indicated by the overlaid black slanted lines in the $y-t$ diagram in panel (f). 
This apparent lateral propagation does not exhibit a simple time-delayed relationship with the photospheric oscillations. 
The propagation can be traced over a lateral distance of approximately 5~Mm from this point of origin.
In contrast, in the weakly magnetized atmosphere, oscillatory patterns extending along the $y$-direction exhibit temporal fluctuations without any clear signature of lateral propagation in the chromosphere. 

\subsection{Intensity}
Figure~\ref{fig2} (g) shows the $y-t$ diagram of $I_{\mathrm{Ca\,II}\, 8542}$ with phase speeds exceeding $7 \ \textrm{km/s}$. 
The occurrence of brightening also exhibits a dependence on the magnetic environment: 
intensity fluctuations exhibit only small amplitudes in the weakly magnetized atmosphere, whereas they are substantially larger in the expanding fluxtube at $y=15^{\prime\prime}$-$28^{\prime\prime}$. 
The brightening is frequently associated with the temporal transition of $v_{\mathrm{Ca\,II}\,8542}$ from redshift to blueshift, particularly in the example marked by the left pointing arrow. 
To facilitate comparison with the chromospheric velocity field, another left-pointing arrow is overlaid at the corresponding coordinates in panel (f). 
Furthermore, the brightening also exhibits diagonal tracks in the $y$–$t$ diagram, similar to those seen in $v_{\mathrm{Ca\,II}\,8542}$. 
In contrast, brightening in the weakly magnetized atmosphere remains predominantly horizontally aligned in the $y-t$ diagram. \\


\subsection{Spectral shape ($\lambda$-t diagram)}
\begin{figure}[h!]
   \centering
   \includegraphics[width=0.9\linewidth]{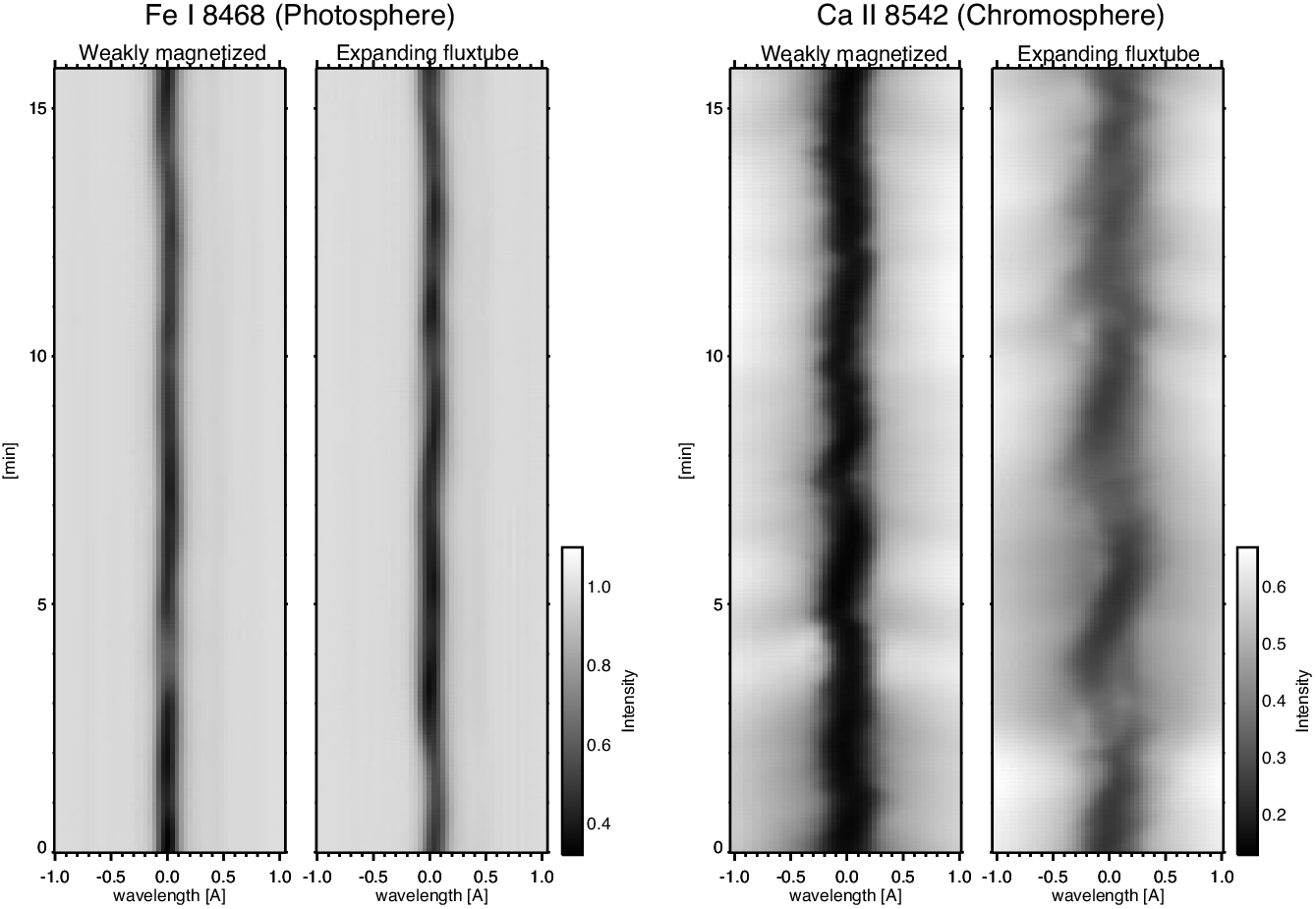}
      \caption{
      $\lambda-t$ diagrams of the photospheric \ion{Fe}{1} 8468\AA~and the chromospheric \ion{Ca}{2} 8542\AA~lines. 
      In each pair, the left panel is at a weakly-magnetized atmosphere ($y=50.5^{\prime\prime}$), while the right panel is at the expanding fluxtube ($y=26.7^{\prime\prime}$). 
      The wavelength abscissae are provided relative to the mean of each spectral line. 
      The grey color scales are shared between each pair. 
\label{fig:cc}}
         \label{fig3}
   \end{figure}

Figure~\ref{fig3} presents two examples of $\lambda -t$ diagrams of Stokes $I$ for the photospheric \ion{Fe}{1} 8468 \AA~line and the chromospheric \ion{Ca}{2} 8542 \AA~line, representing a weakly-magnetized region (at $y=50.5^{\prime\prime}$) and an expanding fluxtube (at $y=26.7^{\prime\prime}$). 
The photospheric \ion{Fe}{1} 8468 \AA~line exhibits only small perturbations in both regions. 
On the other hand, the chromospheric \ion{Ca}{2} 8542 \AA~line in the expanding fluxtube shows significantly large fluctuations and a sawtooth pattern, in which a gradual redshift is followed by an abrupt blueshift. 
Intensity enhancements of the line core coincide with these abrupt transitions from redshift to blueshift, particularly at $t=2$–$3$ and $t=7$–$8$ min. 
Such brightening may result from shock-induced enhancement of the source function:
while the \ion{Ca}{2} 8542 \AA~ line is formed under non-LTE (local thermodynamic equilibrium) condition, the passage of a shock can increase the electron density enough to bring the source function closer to LTE. 
This interpretation is motivated by the explanation originally proposed by \citet{Carlsson1997} for the shock-associated brightening of the \ion{Ca}{2 \ H \ K} lines. 
During shock propagation, the temperature may increase by more than 1~kK, and in some cases by several kK \citep{Mathur2022}. \\ 

\section{Discussion and Summary} \label{sec:highlight}
Our study provides the first observational link between chromospheric magnetic fields and acoustic-wave propagation, based on spectropolarimetric measurements at photospheric and chromospheric heights. 
The observations show that magnetoacoustic portals are associated with fluxtubes that expand toward the chromosphere, thereby producing the inclined magnetic geometry at the tube periphery required to reduce the acoustic cutoff frequency. 
In addition to fluxtube expansion, an overall inclination of the tube axis would also contribute to forming the magnetoacoustic portal, possibly associated with the footpoint of a magnetic loop \citep{Martinez2010, Wiegelmann2010}. 
Such an inclination may be observed around $y=27^{\prime\prime}$, where the centroid position shifts between the photosphere and chromosphere. 
However, the evidence remains inconclusive, and further statistical analysis is required to clarify this possibility.  \\ \indent
It should be noted that not all photospheric fluxtubes act as magnetoacoustic portals. 
For example, the fluxtube located at $y=38.5^{\prime\prime}$ shows evidence of expansion; 
nevertheless, the chromospheric velocity pattern above it exhibits neither strong amplitudes nor low-frequency variations.
Its measured photospheric field strength is only $\sim$100 $\mathrm{G}$, well below the equipartition value ($\sim$400 G; \citealt{Lin1995}), suggesting that the field strength may be insufficient to produce a low plasma-$\beta$ environment. 
However, WFA-retrieved LOS magnetic fields assume a filling factor of unity and are affected by the field inclination, potentially underestimating the intrinsic field strength. 
This limitation should be taken into account, particularly for fluxtubes that exhibit signatures of magnetoacoustic portals despite measured photospheric field strengths of only 200--300 G (below the equipartition value). 
This may indicate that the magnetic filling factor is significantly below unity, implying the presence of unresolved magnetic elements \textcolor{red}{\citep{Stenflo2011, Riethmuller2014}} smaller than the SCIP spatial resolution ($\sim$150 km). 
In such cases, the intrinsic field strength could locally exceed the equipartition value.
More remarkably, a magnetoacoustic portal signature is detected in a magnetic feature with a measured photospheric field strength of only 43 G. 
In SDO/HMI, the same feature would correspond to only about 20 G and could even be barely detectable. 
This result highlights the importance of extending statistical investigations of magnetoacoustic portals to small magnetic elements, such as internetwork features. 
The current dataset is, however, too limited to draw statistical conclusions. 
Only eight photospheric magnetic features were identified, five of which extend into the chromosphere and four of which exhibit signatures of magnetoacoustic portals. 
Previous estimates of the acoustic energy flux transported through magnetoacoustic portals in quiet regions \citep{Jefferies2006, Rajaguru2019} relied on SOHO/MDI \citep{Scherrer1995} and SDO/HMI observations, which cannot reliably identify the weak internetwork magnetic features. 
The contribution of such weak magnetic features may have been underestimated, implying that the acoustic energy flux transported through magnetoacoustic portals could be larger than previously estimated. \\ \indent
In the chromosphere, acoustic oscillations exhibit apparent lateral propagation within the expanding fluxtube. 
Moreover, the propagation pattern does not show a simple correspondence with the photospheric velocity field, consistent with previous observations \citep{Lites1993, Vecchio2009, Kumar2023}. 
These results support a scenario in which slow-mode waves propagate upward through the magnetoacoustic portal and are subsequently guided by the expanding magnetic geometry in the chromosphere, as demonstrated by numerical simulations \citep{Bogdan2003, Heggland2011}. 
The observed propagation distance of approximately 5~Mm suggests that the influence of a magnetoacoustic portal is not confined to the immediate vicinity of the fluxtube but can extend over a substantially larger chromospheric area.  \\ \indent
Our results reveal clear sawtooth patterns in the \ion{Ca}{2} 8542\AA~line within expanding fluxtubes, whereas no such signatures are seen in the weakly magnetized atmosphere. 
Similar chromospheric shock signatures in quiet regions have been reported \citep{Vecchio2009,Cauzzi2009}, although those studies lacked spectropolarimetric measurements. 
Our observations suggest that expanding fluxtubes offer favorable conditions for the upward propagation of low-frequency waves, thereby increasing chromospheric velocity amplitudes. 
As these amplitudes approach the local sound speed, nonlinear wave steepening becomes more efficient, promoting shock formation.
In addition, the efficiency of shock formation also depends on the amplitude of the photospheric velocity fluctuations. 
This is illustrated by the example marked by the left-pointing arrow drawn in Fig.~\ref{fig2}, where enhanced photospheric oscillations coincide with strong chromospheric intensity brightening. 
Such behavior is consistent with previous observations \citep{Vecchio2009, DePontieu2004}. 
Thus, at the chromospheric heights probed by the \ion{Ca}{2} 8542\AA~line, approximately 1.0--1.5 Mm \citep{Quintero_Noda2017b}, shock signatures are not ubiquitous but are preferentially created within expanding fluxtubes. \\ \indent
In summary, this study provides observational evidence that magnetoacoustic portals play an important role in transporting low-frequency acoustic waves into the chromosphere, thereby contributing to chromospheric dynamics through nonlinear wave steepening and shock formation. 
Future statistical investigations of both network and internetwork magnetic fields will be essential for quantifying the acoustic energy flux transported through magnetoacoustic portals and reassessing its contribution to chromospheric heating. 
Furthermore, spectropolarimetric inversions of the SCIP observations \citep{RuizCobo2022} will provide direct measurements of the chromospheric magnetic field inclination and, together with scanned observations, enable reconstruction of the three-dimensional magnetic topology. 
In addition, future improvements in the calibration of the linear polarization measurements may allow the inclination of chromospheric fluxtubes to be constrained.
Such measurements will enable a quantitative assessment of how magnetic geometry modifies the acoustic cutoff frequency and regulates chromospheric wave propagation and shock formation.

\begin{acknowledgments}
{\sc Sunrise~iii} is supported by funding from the Max-Planck-F\"orderstiftung (Max Planck Foundation), NASA under Grants \#80NSSC18K0934 and \#80NSSC24M0024 (``Heliophysics Low Cost Access to Space'' program), and the ISAS/JAXA Small Mission-of-Opportunity program, as well as JSPS KAKENHI Grant Numbers JP18H05234 and JP23K25916. 
This research has received financial support from the European Union's Horizon~2020 research and innovation program under grant agreement No.~824135 (SOLARNET) and No.~101097844 (WINSUN) from the European Research Council (ERC). 
It has also been funded by the Deutsches Zentrum f\"ur Luft- und Raumfahrt~e.V.\ (DLR, grant No.~50~OO~1608). 
The Spanish contributions have been funded by the Spanish MCIN/AEI under projects RTI2018-096886-B-C5 and PID2021-125325OB-C5, and by the ``Center of Excellence Severo Ochoa'' awards to IAA-CSIC (SEV-2017-0709, CEX2021-001131-S), all co-funded by European ERDF funds, ``A way of making Europe''.
The research activities and the flight operation of the SCIP team members, R. T. Ishikawa, M. Kubo, Y. Kawabata, and T. Oba, have been supported by JSPS KAKENHI Grant Numbers JP23KJ0299, JP24K07105, JP23K13152, and JP21K13972, respectively. 
This work was carried out by the joint research program of the Institute for Space-Earth Environmental Research, Nagoya University. 
CQN acknowledges support from Grants PID2022-136563NB-I00/10.13039/501100011033, and PID2024-156538NB-I00 and PID2024-156066OB-C55 funded by MCIN/AEI/10.13039/501100011033. 
\end{acknowledgments}

\bibliographystyle{aasjournalv7}
\bibliography{sola_bibliography_example.bib}



\end{document}